\documentclass[%
     reprint,
      amsmath,amssymb,
      aps,
      pra,
]{revtex4-2}

\usepackage[english]{babel}

\usepackage{comment}
\usepackage{graphicx}
\usepackage{dcolumn}
\usepackage{bm}
\usepackage{hyperref}
\usepackage{amsmath} 
\usepackage{lipsum}
\usepackage{float}
\usepackage{blindtext}
\usepackage{amsmath,amsfonts,amsthm,physics}
\usepackage{adjustbox}
\usepackage{comment}                            

\newcommand{\id}{\mathbb{I}}

\newcommand{\e}{\varepsilon}

\newtheorem{lemma}{Lemma}

\begin{document}

\preprint{APS/123-QED}

\title{Generalised simultaneous transmission of arbitrary quantum states and classical information}

\author{Nicholas Zaunders}
\email{n.zaunders@uq.edu.au}
\affiliation{
    School of Mathematics and Physics, University of Queensland, St Lucia, Queensland 4072, Australia.
}

\author{Timothy C. Ralph}
\affiliation{
    School of Mathematics and Physics, University of Queensland, St Lucia, Queensland 4072, Australia.
}

\begin{abstract}
We present a protocol which allows for arbitrary optical quantum states to simultaneously carry and transmit classical data, without sacrificing the integrity of either the quantum or classical information. Our scheme encodes classical information via displacements in the phase space prior to transmission and retrieves each classical symbol via a Gaussian continuous-variable teleportation. The original quantum state is then restored by guessing the the original displacement and performing the appropriate inverse operation. In the limit of sufficiently high classical signal and high squeezing, we show that our scheme is capable of perfectly reconstructing both the input classical signal and the input quantum state without loss of coherence. An example is given in terms of the transmission of a dual-rail Bell state.
\end{abstract}

\maketitle

\section{Introduction}
Quantum communication using optical-frequency quantum states promises a breadth of information-processing abilities not accessible to ordinary classical information theory \cite{nielsen_quantum_2010}. Moreover, a key feature of quantum communications protocols are their shared qualities - both physically and information-theoretically - with existing classical telecommunication theory. It is therefore a problem of both fundamental \cite{bennett_quantum_2004} as well as applied interest to determine whether and to what degree quantum and classical information can coexist within a single channel. This problem was first addressed in the general case by Devetak and Shor \cite{devetak_capacity_2005} and subsequently in \cite{bradler_trade-off_2010, hsieh_entanglement-assisted_2010, min-hsiu_hsieh_trading_2010}; later, a trade-off coding scheme proposed by Wilde et al. \cite{wilde_information_2012} encoding classical information in permutations of error-corrected quantum states showed that \textit{simultaneous quantum-classical communication} (SQCC) schemes could achieve significant advantage over the bosonic channels relevant to practical telecommunications, compared to equivalent non-simultaneous schemes.

Later, an explicit protocol for the practical simultaneous transmission of quantum and classical information was proposed \cite{qi_simultaneous_2016, qi_noise_2018}, where the classical signal is appended onto a quantum protocol by directly modulating the position of the quantum states in the phase space. This allows the receiver to obtain classical and quantum information with a single measurement, though at the cost of a reduction in the effective classical signal-to-noise as well as coherence of the quantum state \cite{kumar_experimental_2019, zaunders_enhanced_2025}. However, a limitation of this particular simultaneous communication scheme is that the measurement required to extract the classical data also destroys the quantum state; hence the scheme is only compatible with continuous-variable quantum key distribution (CV-QKD) schemes in point-to-point networks.

In this work, we propose a protocol which supports the practical transmission of completely arbitrary quantum states and classical symbols, over arbitrary network configurations, using continuous-variable teleportation \cite{braunstein_teleportation_1998, braunstein_criteria_2000, braunstein_quantum_2005}, which we call classically-modulated quantum communication (CMQC). Our scheme encodes classical data by modulating the mean position of a quantum state in the phase space, as in the simultaneous quantum-classical communication (SQCC) of \cite{qi_simultaneous_2016}, where each position corresponds to a classical symbol. Our scheme differs from ref. \cite{qi_simultaneous_2016} in that upon reception, the displaced quantum state is passed through a continuous-variable teleportation circuit; the teleportation measurements allows the classical symbol to be estimated directly while the quantum state is propagated via a unitary operation on the teleported mode. Knowing the classical displacement then allows the inverse displacement to be applied to the teleported state, returning it to the original state. In the limit of high squeezing and high classical signal, the protocol allows any arbitrary input state to be perfectly reconstructed while also allowing for the simultaneous transmission of classical data without additional modes or channel uses. Our scheme therefore achieves greater generality than trade-off encoding-based protocols while also retaining the simplicity of continuous-variable SQCC \cite{kumar_experimental_2019}. We describe the protocol first in the general case and subsequently provide an example of the transmission of a dual-rail Bell state.

\section{The CMQC protocol}
\begin{figure}
    \centering
    \includegraphics[width=\columnwidth]{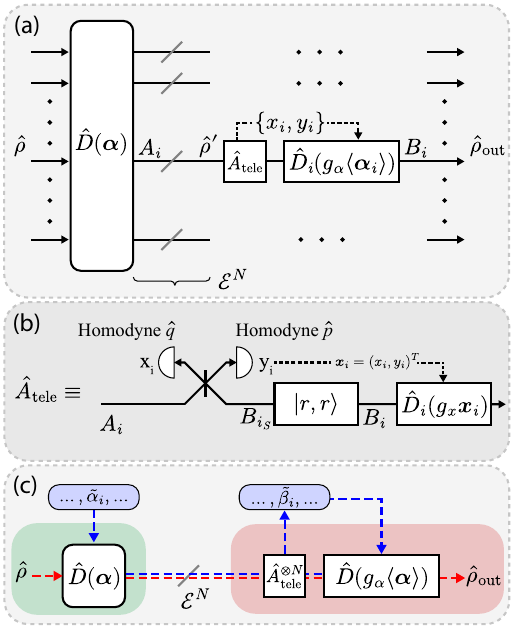}
    \caption{Diagram showing the CMQC protocol (a) as well as the teleportation circuit $\hat A_\mathrm{tele}$ (b). The $N$-mode state $\hat \rho$ Alice wishes to transmit is passed through the displacement operator $\hat D(\bm{\alpha})$, which encodes $N$ classical symbols by displacing each arm in the phase space according to some coherent-state classical protocol. The displaced state is then transmitted over the $N$-mode channel $\mathcal{E}^N$ to Bob. Bob teleports each mode of the displaced transmitted state $\hat \rho'$ by mixing the incoming mode with one half of an EPR state of squeezing $r$ on a balanced beamsplitter, then performing a dual homodyne measurement and correcting the output state via the measurement-dependent unitary operation $\hat D_i(g_x\bm{x}_i)$. The measurement outcomes $\bm{x}_i = (x_i,y_i)^T$ of each arm allow Bob to probabilistically infer the classical displacement and consequently Alice's data stream on a shot-by-shot basis. Lastly, Bob removes the classical displacement from the teleported state by estimating the correct inverse displacement operation $\hat D_i(g_\alpha \langle \bm{\alpha}_i \rangle)$ based on the measurement outcomes. (In practical operation, both displacement unitaries would be combined into a single operation.) (c) Diagram tracking the flow of classical information (blue) and quantum information (red) through the stages of the protocol.}
    \label{fig:protocol_diagram}
\end{figure}

Figure \ref{fig:protocol_diagram} shows the structure of the CMQC protocol. Alice wishes to send some arbitrary $N$-mode state $\hat \rho$, which may be part of some larger entangled state, to Bob, as well as some classical information which she encodes onto $\hat \rho$ by performing the unitary phase-space displacement
\begin{align}
    \hat D(\bm{\alpha}) \equiv \bigotimes_{i = 1}^N \hat D_i(\bm{\alpha_i})
\end{align}
i.e. each mode $A_i$ is independently displaced by the vector displacement $\bm{\alpha}_i \in \mathbb{R}^2$. The displacement on each mode may be arbitrary, but for convenience we assume is chosen from some coherent-state classical encoding, such as phase-shift-keying (PSK), such that each mode encodes a single complex symbol $\tilde \alpha_i$ belonging to some classical alphabet $\{\tilde \alpha_1, \dots, \tilde \alpha_{m} \}$ of size $m = 2^\ell$. Each symbol encodes $\ell$ bits such that the length of the message is $\ell N$. The displacement $\bm{\alpha}_i$ depends on the classical symbol according to $\bm{\alpha}_i~=~(\mathrm{Re}[\tilde \alpha_i], \mathrm{Im}[\tilde \alpha_i])^T$. The displaced state is then transmitted to Bob over some arbitrary $N$-mode channel $\mathcal{E}^N$ such that the output state is given by $\hat \rho' = \mathcal{E}^N\left[ \hat D(\bm{\alpha}) \hat \rho \hat D^\dagger(\bm{\alpha}) \right]$.

In order to extract the classical information contained in $\hat \rho'$ without destroying it, Bob performs a continuous-variable teleportation \cite{braunstein_teleportation_1998, braunstein_criteria_2000, braunstein_quantum_2005} on each received mode (Fig. \ref{fig:protocol_diagram}.b). He first mixes the incoming mode $A_i$ with one mode $B_{i_S}$ of an EPR state $\hat \rho_i^\mathrm{EPR} = \ket{r,r}\bra{r,r}_{B_{i_s}B_i}$ of squeezing $r$ on a balanced beamsplitter $\hat U_\mathrm{mix}$. He then performs a dual homodyne measurement $\hat \Pi_\mathrm{tele}$ on each mode, corresponding to homodyne measurement of the $\hat p$ quadrature of the mode $A_i$ and homodyne measurement of the $\hat q$ quadrature of the mode $B_{i_S}$, from which he obtains the measurement results $y_i$ and $x_i$ respectively. 

The outcomes $\bm{x} = (x_1,y_1,\dots)^T$ are distributed according to the norm of the state post-measurement
\begin{align}
    P(\bm x) &= \Tr[ \hat \Pi_\mathrm{tele} \hat U_\mathrm{mix}^{\otimes N} \left( \hat \rho' \otimes \bigotimes_{i = 1}^N \hat \rho_i^\mathrm{EPR} \right) \hat U_\mathrm{mix}^{\otimes N}{}^\dagger].
\end{align}
Because dual homodyne measurement is a tomographic measurement of the phase-space distribution of the function \cite{lvovsky_continuous-variable_2009}, the outcomes $\bm{x}_i = (x_i, y_i)^T$ measured on each mode will necessarily be distributed around the central classical displacement $\bm{\alpha_i}$ (assuming that the classical displacement is large in comparison to the mean value of $\hat \rho$ prior to displacement). $\bm{x}_i$ will therefore be a good predictor of the classical symbol $\tilde \alpha_i$, and so Bob records his inferred classical symbol $\tilde \beta_i$ on each mode $B_i$ via
\begin{align}
    \tilde \beta_i &= \begin{cases}
        \tilde \alpha_1 & x_1,y_1 \in \Omega_1 \\
        &\vdots \\
        \tilde \alpha_i & x_i,y_i \in \Omega_i \\
        &\vdots \\
        \tilde \alpha_{m} & x_m,y_m \in \Omega_{m}. \\
    \end{cases}
\end{align}
Here $\Omega_i$ corresponds to the regions of the phase space that are associated with each classical symbol in the classical encoding. The likelihood $P(\tilde \beta_i \neq \tilde \alpha_i)$ that Bob records an incorrect symbol on any mode is simply given by the probability that the measurement result for the mode lies outside the correct region; this is called the symbol-error rate $e_S$ and is given by
\begin{align}
    e_S &= P(x_i,y_i \notin \Omega_i) = 1 - \int_{\Omega_i} \dd^{2N}\bm{x} \ P(\bm{x}).
\end{align}

After the classical message is obtained, Bob then uses his knowledge of the measurement results to teleport $\hat \rho'$ onto the output modes $B_i$. For each mode, he first performs the unitary displacement
\begin{align}
    \hat D_i(g_x \bm{x}_i) &= \hat D_{B_i}[(g_x x_i, g_x y_i)^T]
\end{align}
on the unmeasured mode $B_{i}$ of the EPR state $\ket{r,r}$.

The output of the state post-teleportation thus approximates the state $\hat \rho'$. The teleported state is therefore also displaced in the phase space by the original amount $\bm{\alpha}_i$, less the effect of any attenuation introduced by the channel or teleporter. However, our goal is not to recover $\hat \rho'$ but rather to approximate the state $\mathcal{E}^N(\hat \rho)$, i.e. the equivalent channel output of the original non-displaced state. The final step of the protocol thus undoes the effect of Alice's classical displacement by re-displacing each mode in the inverse direction. 

Unlike the teleportation unitary $\hat D_i(g_x \bm{x}_i)$, however, Bob is not able to generate the correct displacement operation $\hat D(-g_\alpha \bm{\alpha})$, since he cannot directly access the correct classical multimode displacement $\bm{\alpha}$ as he can $\bm{x}$: he can only infer them probabilistically based on the outcome of the teleportation. Instead he is limited to performing the non-Gaussian operation $\hat D_i(-g_\alpha \langle \bm{\alpha}_i \rangle)$ on each mode, where
\begin{align}
    \langle \alpha_i\rangle &= (\mathrm{Re}[\tilde \beta_i], \mathrm{Im}[\tilde \beta_i])^T.
\end{align}
Bob's strategy is to assume the measurement correctly reflects the classical displacement, and hence the appropriate inverse displacement is the inverse of the inferred displacement. Symbol errors, which occur when noise associated with the quantum state or the teleporter cause the measured values to deviate significantly from the mean displacement, lead Bob to perform the incorrect re-displacement on the mode. The total output state $\hat \rho_\mathrm{out}$ is therefore composed predominantly of a state approximating $\mathcal{E}^N(\hat \rho)$, i.e. the output state where the inverse displacement operation is successful, with the size of the contribution increasing as $e_S$ decreases; the rest of the state is composed of `impurities' comprised of contributions from states approximating $\mathcal{E}^N(\hat \rho)$ but with non-zero displacement in the phase space.

In total, the protocol is equivalent to the CPTP map $\hat \rho~\longrightarrow~\hat \rho_\mathrm{out}~=~\mathcal{E}_\mathrm{CMQC}(\hat \rho)$ on the arbitrary input state $\hat \rho$, where the channel $\mathcal{E}_\mathrm{CMQC}$ is defined explicitly by
\begin{align}
    &\mathcal{E}_\mathrm{CMQC}(\hat \rho) \label{eq:cmqc_channel}\\
    &= \hat D(-g_\alpha \langle \bm{\alpha} \rangle) \hat A_\mathrm{tele} \mathcal{E}^N \left[ \hat D(\bm{\alpha}) \hat \rho_\mathrm{in} \hat D^\dagger(\bm{\alpha}) \right] \hat A_\mathrm{tele}^\dagger \hat D^\dagger(-g_\alpha \langle \bm{\alpha} \rangle) \notag
\end{align}
for $\hat \rho_\mathrm{in}$ the total input product state comprising Alice's input state $\hat \rho$ and Bob's $N$ EPR resources $\hat \rho_i^\mathrm{EPR}$. Here $\hat A_\mathrm{tele}$ comprises the $N$ separate teleportation circuits, where each $i^\mathrm{th}$ input hybrid mode is mixed with an EPR state, measured, and the outgoing mode fed-forward via $\hat D_i(g_x \bm{x}_i)$.

Making some simplifications to Eq. \eqref{eq:cmqc_channel} yields some insights into the behaviour of $\mathcal{E}_\mathrm{CMQC}$. Firstly, let us assume that the channel $\mathcal{E}^N$ is equivalent to $N$ independent and identical pure-loss channels $\Lambda[\eta]$ of transmissivity $\eta$, such that $\mathcal{E}^N \equiv \Lambda^{\otimes N}[\eta]$. We note that Alice's initial classical displacement operation $\hat D(\bm{\alpha})$ now becomes equivalent to an attenuated displacement $\hat D(\sqrt{\eta}\bm{\alpha})$ performed after the channel. Secondly, we can exploit the fact that the teleportation circuit possesses a degree of freedom in the choice of the electronic gain $g_x$ applied to the feedforward operation $\hat D_i(g_x \bm{x}_i)$; choosing $g_x = -\sqrt{2} \tanh r$ allows us to model the total effect of each teleportation as equivalent to a secondary pure-loss channel $\Lambda[\tau]$ of transmissivity $\tau~=~\tanh^2 r$ \cite{ralph_characterizing_1999}. Hence
\begin{widetext}
    \begin{align}
        \mathcal{E}_\mathrm{CMQC}(\hat \rho) &= \hat D(-g_\alpha \langle \bm{\alpha} \rangle) \hat A_\mathrm{tele} \hat D(\sqrt{\eta}\bm{\alpha}) \Lambda^{\otimes N}[\eta]\left( \hat \rho_\mathrm{in} \right) \hat D^\dagger(\sqrt{\eta}\bm{\alpha}) \hat A_\mathrm{tele}^\dagger \hat D(-g_\alpha \langle \bm{\alpha} \rangle)^\dagger. \label{eq:cmqc_channel_effective_1}\\
        &= \hat D(-g_\alpha \langle \bm{\alpha} \rangle) \Lambda^{\otimes N}[\tau] \left( \hat D(\sqrt{\eta}\bm{\alpha}) \Lambda^{\otimes N}(\eta)\left[ \hat \rho \right] \hat D^\dagger(\sqrt{\eta}\bm{\alpha}) \right) \hat D^\dagger(-g_\alpha \langle \bm{\alpha} \rangle) \\
        &= \hat D(\sqrt{\tau\eta} \bm{\alpha} - g_\alpha \langle \bm{\alpha} \rangle) \Lambda^{\otimes N}[\tau\eta] \left( \hat \rho \right) \hat D^\dagger(\sqrt{\tau \eta} \bm{\alpha} - g_\alpha \langle \bm{\alpha} \rangle). \label{eq:cmqc_channel_effective}
    \end{align}
\end{widetext}
The total effect of the channel on Alice's quantum signal is therefore equivalent to an initial attenuation by a factor $\tau \eta$, followed by a random phase-space displacement $\sqrt{\tau \eta} \bm{\alpha}-g_\alpha \langle\bm{\alpha}\rangle$, which encodes the non-Gaussian noise contribution arising from the corrected classical signal.

Describing the channel in this way allows us to consider two important limiting cases for the CMQC protocol. In the first, we consider the limiting case of high classical signal-to-noise (SNR), where the likelihood of symbol errors vanishes, or equivalently Bob can perfectly predict the classical displacement of the state from his measurement results $\bm{x}_i$ (assuming the channel $\mathcal{E}^N$ is known). Formally, Bob's estimated displacement $\langle \bm{\alpha} \rangle$ approaches the correct value with unity probability:
\begin{align}
    \beta_i \longrightarrow \alpha_i \Longrightarrow-g_\alpha \langle \bm{\alpha} \rangle \longrightarrow -g_\alpha \bm{\alpha}.
\end{align}
Thus Bob can perfectly remove Alice's displacement from the state. In this case, for the appropriate choice of $g_\alpha$, the channel $\mathcal{E}_\mathrm{CMQC}$ approaches
\begin{align}
    \mathcal{E}_\mathrm{CMQC}(\hat \rho) \longrightarrow \Lambda^{\otimes N}(\tau \eta) \left[ \hat \rho \right] \label{eq:cmqc_channel_limA}
\end{align}
Equivalently, if $\mathcal{E}^N$ is not a lossy bosonic channel, for most cases $\mathcal{E}_\mathrm{CMQC}$ will approach $\mathcal{E}^N$ (see Appendix).

In the second limiting case, we consider the limit of teleportation with an infinitely squeezed EPR state. It is a known result \cite{braunstein_teleportation_1998, pirandola_quantum_2006} that in the limit of infinite squeezing CV teleportation can perfectly reconstruct the input state; from Eq. \eqref{eq:cmqc_channel_effective} we find
\begin{align}
    &\mathcal{E}_\mathrm{CMQC}(\hat \rho) \label{eq:cmqc_channel_limR} \\
    &\longrightarrow \hat D(\sqrt{\eta} \bm{\alpha} - g_\alpha \langle \bm{\alpha} \rangle) \mathcal{E}^{\otimes N}[\eta] \left( \hat \rho \right) \hat D^\dagger(\sqrt{\eta} \bm{\alpha} - g_\alpha \langle \bm{\alpha} \rangle) \notag
\end{align}
Importantly, taking the further limit of Eq. \eqref{eq:cmqc_channel_limA} over high squeezing is \textit{not} equivalent to taking the further limit of Eq. \eqref{eq:cmqc_channel_limR} over high classical signal. This asymmetry arises from the reverse coupling of the quantum channel onto the classical: as the teleportation squeezing increases, the effective noise on the classical signal also grows as $\cosh 2r$. Hence even in the best-case scenario of a quantum state and channel with minimal additional noise, the classical signal must grow at least exponentially with the degree of squeezing to maintain a workable signal-to-noise ratio. In the example of a pure-loss channel and a PSK encoding, the signal-to-noise ratio for a simple encoding is approximately
\begin{align}
    \mathrm{SNR} &\sim \frac{\sqrt{\eta} \abs{\bm{\alpha}_i}}{\cosh2r}
\end{align}
and so we require a classical signal strength on the order of $\abs{\bm{\alpha}_i} \sim \eta^{-1/2} \cosh 2r$.
When the above condition is not met, the frequency of classical errors increases sharply. This causes the effective displacement operator $\hat D(\sqrt{\tau \eta} \bm{\alpha} - g_\alpha \langle \bm{\alpha} \rangle)$ in Eq. \eqref{eq:cmqc_channel_effective} to deviate from identity, generating an additional source of uncorrelated noise on the quantum signal. Because of this effect, increasing the squeezing without also increasing the classical signal strength can counterintuitively degrade the quantum signal. A useful corollary of this, however, is that good operation of the protocol does not in general require high levels of squeezing.

 Lastly, the above results describe protocol for a single use of the quantum-classical channel. In practical use, however, the classical symbol is drawn effectively at random each round with probability $2^{-\ell N}$. We are therefore required to model the displaced input state not as a single quantum state corresponding to a single symbol, $\hat D(\bm{\alpha_j}) \hat \rho \hat D^\dagger(\bm{\alpha_j})$, but instead as a stochastic preparation of each possible displaced state corresponding to a possible symbol:
\begin{align}
    \hat \rho \longrightarrow \sum_{j=1}^{2^{\ell N}} 2^{-\ell N} \hat D(\bm{\alpha_j}) \hat \rho \hat D^\dagger(\bm{\alpha_j})
\end{align}
where $\bm{\alpha_j} \in \mathbb{R}^{2N}$ is the phase-space displacement corresponding to the arbitrary collection of $N$ symbols $\{\dots,\tilde \alpha_i,\dots \}$ sent by Alice on any given round. Fortunately, the linearity of the protocol allows us to compute the $\hat \rho_\mathrm{out}$ for each given displacement and sum the result to obtain the full output state. Lastly, assuming each use of the protocol is independently and identically distributed, the output state after many rounds is given by the ensemble state $\hat \rho_\mathrm{ens}$
\begin{align}
    \hat \rho_\mathrm{ens} = \int \dd^{2N} \bm{x} \ P(\bm{x}) \hat \rho_\mathrm{out}(\bm{x}).
\end{align}

\section{Application of the protocol to transmission of a dual-rail Bell state}As a practical and immediately relevant application of our protocol we consider the distribution of Bell states between distant parties. This is currently achieved over significant distances via fibre \cite{neumann_continuous_2022} and free space channels \cite{yin_satellite-based_2017} and is key to cryptographic \cite{brunner_bell_2014} and future quantum repeater \cite{azuma_quantum_2023} protocols. The Bell statistics of an ensemble of states is a rigorous measure of the quality of entanglement. In current experiments dual-rail encodings such as polarisation or time-bins are used. These encodings are loss tolerant in the sense that the Bell statistics of an ensemble can be postselected with respect to valid measurement outcomes to exclude channel or state preparation loss. An ensemble is said to be Bell-violating \footnote{
    Postselected Bell tests are not loophole-free and so do not strictly exclude local models \cite{pearle_hidden-variable_1970, gisin_local_1999}, but since this is not our aim here they are nevertheless a suitable metric for evaluating the protocol.
} for $\abs{S_p} > 2$ \cite{branciard_detection_2011}, where $S_p = S/p_s$ for $S$ the detection-loophole-free CHSH statistic \cite{bell_einstein_1964, clauser_proposed_1969} and $p_s$ the probability of obtaining a valid measurement outcome.

In this example we show that the CMQC protocol can be used to transmit Bell states of arbitrary quality simultaneously with classical information by computing the Bell criterion. We consider the transmission of the dual-rail singlet Bell state $\hat \rho~=~\ket{\phi^-}\bra{\phi^-}$, i.e.
\begin{align}
    \ket{\phi^-} &= \frac{\ket{0_{L_1}}\ket{1}_{L_2} - \ket{1_{L_1}}\ket{0_{L_2}}}{\sqrt{2}} \\
    &\equiv \frac{\ket{1001}_{A_1 A_2 A_{1_s} A_{2_s}} - \ket{0110}_{A_1 A_2 A_{1_s} A_{2_s}}}{\sqrt{2}},
\end{align}
where modes $A_{1,2}$ are retained by Alice and modes $A_{1_S,2_S}$ are transmitted over the channel to Bob, i.e. $N = 2$. For the classical encoding we consider a dual binary phase-shift-keyed (BPSK) protocol on each arm; Alice encodes the classical symbols $\tilde \alpha_{1,2} = \pm \alpha$ on mode $A_{1_S}$ and $\tilde \alpha_{1,2} = \pm i\alpha$ on mode $A_{2_S}$. The signal strength is given by the magnitude of the displacement $\alpha \in \mathbb{R}$. We also assume the channel $\mathcal{E}$ is equivalent to an ensemble of $N$ identical pure-loss channels, since this is a reasonable assumption to make for classical and quantum channels \cite{weedbrook_gaussian_2012} as well as for dual-rail states (which frequently exploit the same physical channel for both rails \cite{kok_linear_2007,lukens_frequency-encoded_2017}). Each channel is characterised by a transmissivity $\eta$ such that $\mathcal{E}^N~\equiv~\Lambda^{\otimes N}(\eta)$.
The full computation of the output state $\hat \rho_\mathrm{ens}$ is performed analytically in the Wigner basis and can be found in Appendix \ref{app:appendix}.

\begin{figure}
    \centering
    \includegraphics[width=\columnwidth]{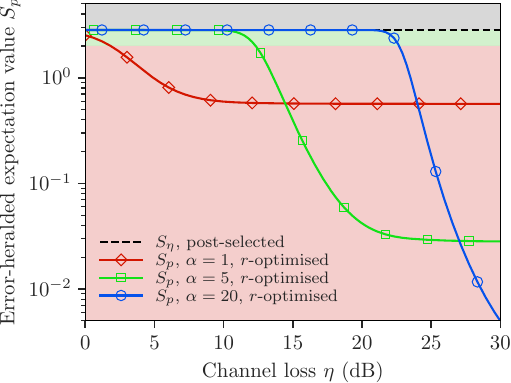}
    \caption{CHSH criterion $S_p$ as a function of channel loss $\eta$ and classical signal strength $\alpha$, evaluated on $\rho_{ens}$. At each point, $S_p$ is optimised over Bob's EPR resource squeezing $r$. Post-selecting the correlations based on successful photon counts in principle permits the protocol to saturate the upper bound of $S_p = 2\sqrt{2}$ at arbitrary loss, albeit at a vanishing rate, for sufficiently large signal sizes. However, for finite signal strength $\alpha$ the protocol can only achieve the optimum correlations up to a finite loss. Past this point, the degradation of the classical signal decreases the SNR to a point where classical re-displacement errors destroy the quantum correlations. The rates of each protocol and optimal $r$ values are shown in Appendix \ref{app:appendix}.}
    \label{fig:pubfig_bellstate_postselected}
\end{figure}

For our example of the dual-rail Bell state, in the limit of high classical signal we retrieve
\begin{align}
    \lim_{\alpha \longrightarrow \infty}p_s &= \eta \tanh^2 r = \eta \tau
\end{align}
which is the expected probability of success for the effective pure-loss channel in the limit of no classical errors; the postselected Bell criterion becomes
\begin{align}
    \lim_{\alpha \longrightarrow \infty} S_p &= \frac{2\sqrt{2} \eta \tanh^2 r}{\eta \tanh^2 r} = 2\sqrt{2}.
\end{align}
Figure \ref{fig:pubfig_bellstate_postselected} shows $S_p$ as a function of channel loss and proves that in the postselected regime the CMQC protocol can transmit a logical state capable of optimally violating the Bell inequality at arbitrary loss for sufficiently high classical signal.

We note here that future works could implement this postselected Bell test in a more robust and fault-tolerant way by instead transmitting a multi-mode \textit{logical} Bell state encoded via some quantum error correction code, such as a parity check code \cite{hayes_loss-tolerant_2008, muralidharan_ultrafast_2014, ewert_ultrafast_2016, ewert_ultrafast_2017}, which heralds the presence of physical errors such as photon loss by making a suitable syndrome measurement on the protocol output. Conducting a Bell test using logical qubits therefore allows us to discard events where an error is detected, and so the calculation of the CHSH statistic is computed only with respect to events where the correct photon is successfully detected.

In this case, the probability $p_s$ is given by the probability of the error code heralding no error, or equivalently the likelihood of obtaining a conclusive measurement result, which we define as Bob measuring a single photon in one detector and a vacuum in the other. Postselecting in this way acts to filter out the effect of the channel, since if a photon is lost to the environment via $\mathcal{E}$ it is detected and automatically discluded from the calculation of $S$. In principle, however, the postselected results shown in Figure \ref{fig:pubfig_bellstate_postselected} suitably reflect the error-corrected scheme.

\section{Conclusion}
In this work we have introduced a scheme which allows for the integration of fully arbitrary quantum communication simultaneously with classical data transfer. We describe the theoretical structure of the protocol and show that in the limit of a sufficiently strong classical signal a single channel use can yield both an arbitrarily-low-error classical symbol as well as a completely arbitrary quantum state with coherence limited only by the channel characteristics. We investigate transmission of a Bell state using the scheme and demonstrate near-ideal postselected Bell statistics for modest classical signal strength and squeezing resources. Our scheme provides a substantial advantage over the SQCC scheme of \cite{qi_simultaneous_2016} in that it enables any classical network to simultaneously host a parallel quantum network, or vice versa, with minimal loss of quantum coherence or classical error rate, regardless of the network topology or desired quantum protocol. Our scheme has immediate applicability to future implementations of the quantum internet, which will require high volumes of both classical and quantum data transfer in a way that is low-error and compact; for example, we envision a classical network which performs entanglement distribution simultaneously with data transfer and stores the distributed e-bits in quantum memories placed at each node, which can be distilled and consumed on demand.

\begin{acknowledgements}
The Australian Government supported this research through the Australian Research Council’s Linkage Projects funding scheme (Project No. LP200100601) and the Centre of Excellence for Quantum Computation and Communication Technology (Project No. CE170100012). The views expressed herein are those of the authors and are not necessarily those of the Australian Government or the Australian Research Council.
\end{acknowledgements}

\newpage
\appendix
\begin{widetext}
\section{Transmitting a dual-rail Bell state via the CMQC protocol}
\label{app:appendix}
\textit{Defining the input state $\hat \rho$.}{---}To begin, we define the target state $\hat \rho$ which Alice wishes to distribute to Bob. This state is the logical singlet Bell state $\ket{\phi^-}\bra{\phi^-}$, where
\begin{align}
    \ket{\phi^-} &= \frac{\ket{0}_{L_1}\ket{1}_{L_2} - \ket{1}_{L_1}\ket{0}_{L_2}}{\sqrt{2}}.
\end{align}
The state is encoded in a dual-rail configuration, meaning each qubit is encoded in two modes \cite{kok_linear_2007}. Here we define two pairs of modes $\{A_1,A_{2}\}$, and $\{A_{1_S},A_{2_S}\}$ to represent each qubit such that for each qubit logical 0 corresponds to a photon in the top modes $A_1$, $A_{1_S}$ and logical 1 corresponds to a photon in the bottom modes $A_2$, $A_{2_S}$ (Figure \ref{fig:supfig_bellstate_diagram}).

\begin{figure}[!htb]
    \centering
    \includegraphics{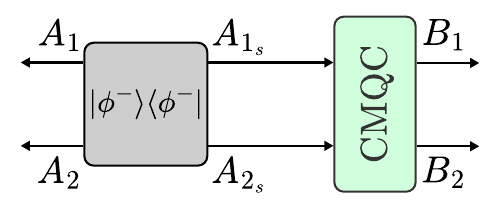}
    \caption{Diagram showing the transmission of a dual-rail Bell state $\ket{\phi^-}\bra{\phi^-}$ via the CMQC protocol. Each half of the dual-rail state is encoded in the modes $A_1, A_{2}$ and $A_{1_s}, A_{2_s}$ respectively, where the logical states are given by $\ket{0}_{L_1} = \ket{1}_{A_1}\ket{0}_{A_2}$, $\ket{1}_{L_1} = \ket{0}_{A_1}\ket{1}_{A_2}$, $\ket{0}_{L_2} = \ket{1}_{A_{1_s}}\ket{0}_{A_{2_s}}$ and $\ket{1}_{L_2} = \ket{0}_{A_{1_s}}\ket{1}_{A_{2_s}}$.The half corresponding to the logical mode $L_2$ is sent by Alice to Bob, where it is teleported onto the modes $B_1, B_2$ such that the total output of the protocol becomes an approximate equivalent dual-rail Bell state over the modes $A_1, A_{2}$ and $B_1, B_{2}$.}
    \label{fig:supfig_bellstate_diagram}
\end{figure}

The physical state Alice wishes to send is therefore
\begin{align}
    \ket{\phi^-} = \frac{\ket{1001}_{A_1 A_2 A_{1_s} A_{2_s}} - \ket{0110}_{A_1 A_2 A_{1_s} A_{2_s}}}{\sqrt{2}} \label{eq:bellstate_physical}
\end{align}
where Alice retains the modes $A_{1,2}$ and transmits the modes $A_{1_s,2_s}$ to Bob via some channel $\mathcal{E}$. We perform our analysis of the protocol in terms of the Wigner representation \cite{wigner_quantum_1932, curtright_concise_2014}. Alice's input state is represented by the Wigner function $W_{\hat \rho}$, where in general for an $N$-mode state
\begin{align}
    W_{\hat \rho}(\bm{q}) &= \frac{1}{(2\pi)^{2N}} \int \dd^{2N}\bm{\xi} \ e^{\dots + iq_i \xi_{iy}-ip_i\xi_{ix}+\dots} \Tr\left[ \hat D(\bm{\alpha}) \hat \rho \right]
\end{align}
for $\bm{q} = (q_1, p_1, \dots, p_N)^T$, $\bm{\xi} = (\xi_{1x}, \xi_{1y}, \dots, \xi_{Ny})^T$, and 
\begin{align}
    \hat D(\bm{\alpha}) &\equiv \bigotimes_{i=1}^N \hat D_i([\xi_{ix}, \xi_{iy}]^T).
\end{align}
The Wigner function of the state $\hat \rho$ is thus
\begin{align}
    W_{\hat \rho}(\bm{q}) = \frac{1}{{32 \pi ^4}}\exp\bigg[&-\frac{1}{2} \left(p_{A_1}^2+p_{A_{1_S}}^2+p_{A_{2}}^2+p_{A_{2_S}}^2+q_{A_1}^2+q_{A_{1_S}}^2+q_{A_{2}}^2+q_{A_{2_S}}^2\right)\bigg] \notag\\
    \bigg[&2-q_{A_{1}}^2-p_{A_{1}}^2-q_{A_{1_s}}^2-p_{A_{1_s}}^2 -q_{A_{2}}^2-p_{A_{2}}^2-q_{A_{2_s}}^2-p_{A_{2_s}}^2+ \notag\\
    &q_{A_{1}}^2 q_{A_{2_s}}^2+q_{A_{1_s}}^2 q_{A_{2}}^2+p_{A_{1}}^2 p_{A_{2_s}}^2+p_{A_{1}}^2 q_{A_{2_s}}^2 +p_{A_{1_s}}^2 q_{A_{2}}^2+p_{A_{1_s}}^2 p_{A_{2}}^2+p_{A_{2_s}}^2 q_{A_{1}}^2+p_{A_{2}}^2 q_{A_{1_s}}^2 -\notag\\
    &2 p_{A_{1}} p_{A_{1_s}} p_{A_{2}} p_{A_{2_s}}-2 p_{A_{1}} p_{A_{1_s}} q_{A_{2}} q_{A_{2_s}}-2 p_{A_{1}} p_{A_{2}} q_{A_{1_s}} q_{A_{2_s}}+2 p_{A_{1}} p_{A_{2_s}} q_{A_{1_s}} q_{A_{2}} +  \notag\\
    &2 p_{A_{1_s}} p_{A_{2}} q_{A_{1}} q_{A_{2_s}}-2 p_{A_{1_s}} p_{A_{2_s}} q_{A_{1}} q_{A_{2}}-2 p_{A_{2}} p_{A_{2_s}} q_{A_{1}} q_{A_{1_s}} -2 q_{A_{1}} q_{A_{1_s}} q_{A_{2}} q_{A_{2_s}}\bigg]
\end{align}
and $\bm{q} = (q_{A_1},p_{A_1},q_{A_{1_s}},p_{A_{1_s}},q_{A_2},p_{A_2},q_{A_{2_s}},p_{A_{2_s}})$.

\vspace{1em}
\textit{Performing the classical encoding.}{---}Next, Alice encodes the classical data onto the state $\hat \rho$ by displacing it in the phase space. For simplicity, we choose an encoding composed of a dual BPSK scheme, where the top mode $A_{1_s}$ contains a BPSK encoding about the $\hat q$ quadrature and the bottom mode $A_{2_s}$ contains a BPSK encoding about the $\hat p$ quadrature. (This is effectively equivalent to a quadrature phase-shift-keyed encoding on a single channel.) For the mode $A_{1_s}$, the classical alphabet is composed of $2^1$ symbols $\{\tilde \alpha_1 = \alpha, \tilde \alpha_2 = -\alpha \}$ for $\alpha \geq 0$ and encode the classical bitstrings $\{0\}$ and $\{1\}$ respectively. For the mode $A_{2_s}$, the classical alphabet is composed of $2^1$ symbols $\{\tilde \alpha_1 = i\alpha, \tilde \alpha_2 = -i\alpha \}$ for $\alpha \geq 0$ and again encode the classical bitstrings $\{0\}$ and $\{1\}$ respectively. Thus, Alice encodes her $\ell N = 2$ bits of classical information by performing the phase-space displacement
\begin{align}
    \hat D(\bm{\alpha}) &= \hat D_{A_{1_s}}([\pm \alpha, 0]^T) \otimes \hat D_{A_{2_s}}([0,\pm\alpha]^T) \\
    &\equiv \hat D([\pm \alpha, 0, 0, \pm \alpha]^T).
\end{align}
Without loss of generality we assume Alice sends the classical string $\{00\}$, such that 
\begin{align}
    \hat D(\bm{\alpha}) &= \hat D([\alpha, 0, 0, \alpha]^T)
\end{align}
since the phase-space symmetry of the classical protocol guarantees that the the expectation value $S$ does not change with respect to the choice of symbol. Because the displacement operation is a Gaussian unitary with an associated affine form acting on the quadratures of the phase space, we can use the following result \cite{weedbrook_gaussian_2012}

\begin{lemma} \label{lem:wignerfunctransform}
    For a Gaussian unitary $\hat U$ with respective affine map acting on the quadrature operators $\bm{\hat{q}}$
    \begin{align}
        (S, d) : \hat{\bm{q}} \longrightarrow S\bm{q} + \bm{\alpha}
    \end{align}
    the unitary transformation acting on a state $\hat \sigma$
    \begin{align}
        \hat \sigma \longrightarrow \hat U \hat \sigma \hat U^\dagger
    \end{align}
    is equivalent to the transform on the Wigner function
    \begin{align}
        W_{\sigma}(\bm{q}) \longrightarrow W_\sigma(S^{-1}\bm{q} - S^{-1}\bm{\alpha}).
    \end{align}
\end{lemma}

The classical encoding operation has the affine representation
\begin{align}
    \bm{q} \longrightarrow \bm{q} + (0,0,2\alpha,0,0,0,0,2\alpha)^T
\end{align}
and so using Lemma \ref{lem:wignerfunctransform} the displaced state $W_{\hat D(\alpha)\hat \rho \hat D^\dagger(\alpha)}$ is given by
\begin{align}
    W_{\hat D(\alpha)\hat \rho \hat D^\dagger(\alpha)}(\bm{q}) = W_{\hat \rho}(q_{A_1},p_{A_1},q_{A_{1_s}},p_{A_{1_s}},q_{A_2}-2\alpha,p_{A_2},q_{A_{2_s}},p_{A_{2_s}}-2\alpha).
\end{align}

\vspace{1em}
\textit{Transmission of the displaced state.}{---}The displaced modes are then transmitted across the channel $\mathcal{E}$ to Bob. We assume for simplicity that the channel associated with each mode is a pure-loss channel $\mathcal{E}(\eta)$ of transmissivity $\eta$, i.e.
\begin{align}
    \mathcal{E} = \mathcal{E}_{A_{1_s}}(\eta) \otimes \mathcal{E}_{A_{2_s}}(\eta)
\end{align}
which is an appropriate model for both quantum and classical communications \cite{weedbrook_gaussian_2012}. To model the channel transmission, we introduce two ancilla vacuum modes, $E_1$ and $E_2$, with the associated Wigner function
\begin{align}
    W_{E_i}(\bm{q}) &= \frac{1}{(2\pi)^2} \exp\left[-\frac{1}{2}\left(q_{E_i}^2+p_{E_i}^2\right)\right].
\end{align}
The Wigner function $W$ of the total product state input to the channel is therefore
\begin{align}
    W &= W_{\hat D(\alpha)\hat \rho \hat D^\dagger(\alpha)} W_{E_1} W_{E_2}.
\end{align}
The channel itself is modelled via the Stinespring dilation \cite{stinespring_positive_1955} with unitary representation $\hat U_\mathrm{chn}$ as a pair of beamsplitters of transmissivity $\eta$ \cite{weedbrook_gaussian_2012} between each transmitted mode and each ancilla mode, with symplectic representation
\begin{align}
    S_\mathrm{chn} &= \begin{pmatrix}
        \id & 0 & \id & 0 & 0 & 0 \\
        0 & \sqrt{\eta} \id & 0 & 0 & -\sqrt{1-\eta} \id & 0 \\
        \id & 0 & \id & 0 & 0 & 0 \\
        0 & 0 & 0 & \sqrt{\eta} \id & 0 & -\sqrt{1-\eta}  \\
        0 & \sqrt{1-\eta} \id & 0 & 0 & \sqrt{\eta} \id & 0 \\
        0 & 0 & 0 & \sqrt{1-\eta}  & 0 & \sqrt{\eta} \id 
    \end{pmatrix}.
\end{align}
The output state $\hat \rho'$ is therefore given by
\begin{align}
    \hat \rho' &= \Tr_{E_1E_2} \left[ \hat U_\mathrm{chn} (\hat D(\bm \alpha)\hat \rho \hat D^\dagger(\bm \alpha) \otimes \hat \rho_{E_1} \otimes \hat \rho_{E_2}) \hat U_\mathrm{chn}^\dagger \right]
\end{align}
with Wigner representation (again using Lemma \ref{lem:wignerfunctransform})
\begin{align}
    W_{\hat \rho'}(\bm{q}) &= \int \dd^4 \bm{q}_{E} \ W_{\hat D(\alpha)\hat \rho \hat D^\dagger(\alpha)}(S_\mathrm{ch}^{-1}\bm{q}) W_{E_1}(S_\mathrm{ch}^{-1}\bm{q}) W_{E_2}(S_\mathrm{ch}^{-1}\bm{q}).
\end{align}

\vspace{1em}
\textit{Teleportation of the received state $\hat \rho'$.}{---}
Recall that $\hat \rho'$ is the state received by Bob and the target of the CV teleportation circuit represented by the operator $\hat A_\mathrm{tele}$ (Figure \ref{fig:protocol_diagram}.b in the main text). The circuit is composed of three steps:
\begin{enumerate}
    \item Each incoming mode $A_{i_s}$ is mixed with one mode $B_{i_s}$ of an EPR state of squeezing $r$ on a balanced beamsplitter.
    \item A dual homodyne measurement is taken with respect to the output ports of the beamsplitter.
    \item A unitary operation is performed on the remaining EPR mode $B_i$ to transfer the correlations and complete the teleportation.
\end{enumerate}
The input joint state to the teleportation circuit is the product state between the input state $\hat \rho'$ and each mode's EPR state:
\begin{align}
    \hat \rho_\mathrm{pre-tele} &= \hat \rho' \otimes \ket{r,r}\bra{r,r}_{B_{1_s} B_1} \otimes \ket{r,r}\bra{r,r}_{B_{2_s} B_2}
\end{align}
where $\ket{r,r}_{AB}$ is the two-mode squeezed vacuum state of squeezing $r$ over modes $A,B$ \cite{weedbrook_gaussian_2012}
\begin{align}
    \ket{r,r}_{AB} &= \mathrm{sech} \ r \sum_{j=0}^\infty \tanh^{j}(r) \ket{j}_A \otimes \ket{j}_B
\end{align}
with Gaussian Wigner function
\begin{align}
    W_\mathrm{AB}^\mathrm{EPR}(q_A,p_A,q_B,p_B) &= \frac{1}{(2\pi)^4} \exp \left[ -\frac{1}{2} \left( q_A^2 + p_A^2 + q_B^2+p_B^2 \right)\cosh2r + (q_A q_B - p_A p_B) \sinh 2r \right].
\end{align}
The total joint state prior to teleportation is thus
\begin{align}
    W_\mathrm{pre-tele}(\bm q) &= W_{\hat \rho'} W_\mathrm{B_{1_s}B_1}^\mathrm{EPR} W_\mathrm{B_{2_s}B_2}^\mathrm{EPR}.
\end{align}
Step 1, the mixing of each incoming mode $A_i$ with the respective EPR mode $B_{i_s}$, is a Gaussian unitary operation $\hat U_\mathrm{mix}$ with symplectic representation
\begin{align}
    S_\mathrm{mix} &= \begin{pmatrix}
        \id & 0 & 0 & 0 & 0 & 0 & 0 & 0 \\
        0 & \frac{1}{\sqrt{2}} \id & 0 & 0 & -\frac{1}{\sqrt{2}} \id & 0 & 0 & 0 \\
        0 & 0 & \id & 0 & 0 & 0 & 0 & 0 \\
        0 & 0 & 0 & \frac{1}{\sqrt{2}} \id & 0 & 0 & -\frac{1}{\sqrt{2}} \id & 0 \\
        0 & \frac{1}{\sqrt{2}} \id & 0 & 0 & \frac{1}{\sqrt{2}} \id & 0 & 0 & 0 \\
        0 & 0 & 0 & 0 & 0 & \id & 0 & 0 \\
        0 & 0 & 0 & \frac{1}{\sqrt{2}} \id & 0 & 0 & \frac{1}{\sqrt{2}} \id & 0 \\
        0 & 0 & 0 & 0 & 0 & 0 & 0 & \id 
    \end{pmatrix};
\end{align}
the state after mixing is given by (Lemma \ref{lem:wignerfunctransform})
\begin{align}
    W_{\mathrm{pre-tele}}(\bm q) \longrightarrow W_{\mathrm{pre-tele}}(S_\mathrm{mix}^{-1} \bm{q}).
\end{align}
Step 2 requires Bob to perform a dual homodyne measurement on the output ports of the mixing beamsplitter in Step 1. More specifically, for each teleportation circuit acting on the $i$th input mode Bob measures the $\hat q$ quadrature of the mixed EPR mode, $q_{B_{i_s}}$, and the $\hat p$ quadrature of the mixed input mode, $p_{A_{i_s}}$, obtaining the measurement results $x_i$ and $y_i$ respectively for each mode. On a mode-by-mode basis this is represented by the projector
\begin{align}
    \hat \Pi_\mathrm{tele} &= \bigotimes_{i=1}^2 \hat \Pi_i,
\end{align}
where
\begin{align}
    \hat \Pi_i &= \id \otimes \ket{\hat p_{A_{i_s}} = y_i}\bra{\hat p_{A_{i_s}} = y_i} \otimes \ket{\hat q_{B_{i_s}} = x_i}\bra{\hat q_{B_{i_s}} = x_i} \otimes \id
\end{align}
where the individual single-mode projectors $\ket{\hat p_{A_{i_s}} = y_i}\bra{\hat p_{A_{i_s}} = y_i}$, $\ket{\hat q_{B_{i_s}} = x_i}\bra{\hat q_{B_{i_s}} = x_i}$ represent projection of the modes $A_i$, $B_{i_s}$ onto the respective momentum and position eigenstates $\ket{y_i}$ and $\ket{x_i}$ \cite{fiurasek_gaussian_2002,eisert_distilling_2002}. The density matrix of the state after the teleportation measurement is \cite{nielsen_quantum_2010}
\begin{align}
    \hat \rho_\mathrm{post-tele} &= \frac{\Tr_{A_{1_s}B_{1_s}A_{2_s}B_{2_s}}\left[ \hat \Pi_\mathrm{tele} \ \hat U_\mathrm{mix} \hat \rho_\mathrm{pre-tele} \hat U_\mathrm{mix}^\dagger \right]}{P(\bm{x})}
\end{align}
for the norm of the state
\begin{align}
    P(\bm{x}) \equiv P(x_1,y_1,x_2,y_2) &= \Tr\left[ \hat \Pi_\mathrm{tele} \left( \hat U_\mathrm{mix} \hat \rho_\mathrm{pre-tele} \hat U_\mathrm{mix}^\dagger \right) \right].
\end{align}
In the Wigner representation measurements are given by the partial integral of the Wigner function weighted by the Wigner representation of the measurement operator \cite{curtright_concise_2014}. It is not difficult to show that the Wigner function of position and momentum eigenstates $\ket{\hat q = x}\bra{\hat q = x}, \ket{\hat p = y}\bra{\hat p = y}$ in the phase space are the delta functions $\delta(q - x)/4\pi, \delta(p - y)/4\pi$; thus
\begin{align}
    W_{\mathrm{post-tele}}(\bm{q}) &= \frac{1}{P(\bm{x})} \ (4\pi)^4 \int \dd^4\bm{q_{A_{i_s}}} \ \dd^4\bm{q_{B_{i_s}}} \ W_{\hat \Pi_\mathrm{tele}}(\bm{q}) W_{\mathrm{pre-tele}}(S_\mathrm{mix}^{-1} \bm{q}) \\
    &= \frac{1}{P(\bm{x})} \int \dd^4\bm{q_{A_{i_s}}} \ \dd^4\bm{q_{B_{i_s}}} \ \delta(p_{A_{1_s}} - y_1)\delta(q_{B_{1_s}} - x_1)\delta(p_{A_{2_s}} - y_2)\delta(q_{B_{2_s}} - x_2) W_{\mathrm{pre-tele}}(S_\mathrm{mix}^{-1} \bm{q}) \\
    P(\bm{x}) &= \int \dd^{16}\bm{q} \ \delta(p_{A_{1_s}} - y_1)\delta(q_{B_{1_s}} - x_1)\delta(p_{A_{2_s}} - y_2)\delta(q_{B_{2_s}} - x_2) W_{\mathrm{pre-tele}}(S_\mathrm{mix}^{-1} \bm{q})
\end{align}

Lastly, Step 3 requires Bob to perform the displacement operation
\begin{align}
    \hat D(g_x \bm{x}) &\equiv \hat D_{B_1}[g_x(x_1,y_1)] \otimes \hat D_{B_2}[g_x(x_2,y_2)^T]
\end{align}
on the output modes $B_i$ of the teleporter. The operation and the gain factor
\begin{align}
    g_x &= -\sqrt{2}\tanh r
\end{align}
are chosen such that the output state is maximally purified \cite{braunstein_quantum_2005}, i.e. the output state is independent of the measurement results $\bm{x} = (x_1,y_1,x_2,y_2)^T$; this corresponds to the teleporter generating an effective pure-loss channel of transmissivity $\tau = \tanh^2r$ \cite{ralph_characterizing_1999}. We also note that in general the state cannot achieve the full purification described by \cite{braunstein_criteria_2000, braunstein_quantum_2005}, since this requires Alice to also perform an equivalent unitary on her modes $A_i$, which requires her to obtain from Bob the measurement outcome $\bm{x}$ via some classical communications channel. Since assuming that a classical channel exists between Alice and Bob undercuts somewhat the purpose of the CMQC protocol, we assume Alice does not have the outcome of the teleportation and so does perform any operation on any part of the entangled state.

The output state of the teleporter is given by
\begin{align}
    \hat \rho_\mathrm{tele} &= \hat D(g_x \bm{x}) \hat \rho_\mathrm{post-tele} \hat D^\dagger(g_x \bm{x})
\end{align}
and so the Wigner representation of the state output by the circuit $\hat A_\mathrm{tele}$ is
\begin{align}
    W_{\mathrm{tele}}(\bm{q}) = W_{\mathrm{post-tele}}(q_{A_1}, p_{A_1}, q_{A_2}, p_{A_2}, q_{B_1} - g_x x_1, p_{B_1} - g_x y_1, q_{B_2} - g_x x_2, p_{B_2} - g_x y_2).
\end{align}

\vspace{1em}
\textit{Decoding the classical data stream and removing the classical displacement.}{---} Because dual homodyne detection is a tomographic process \cite{lvovsky_continuous-variable_2009}, the norm of the state $P(\bm{x})$ indirectly characterises the distribution of the Wigner function in the phase space. While a full tomographic reconstruction is not necessary, Bob can exploit his knowledge of $P(\bm{x})$ along with his measurement results $\bm{x}$ to estimate the mean displacement of the state on a shot-by-shot basis, which he can then use to infer Alice's initial displacement and subsequently the sent classical symbol. In the case of the Bell state, the norm can be explicitly written:
\begin{align}
    &P(\bm{x}) = \frac{\exp \left[-\frac{(x_1 - \sqrt{2}\sqrt{\eta}\alpha)^2 + y_1^2 + x_2^2 + (y_2 - \sqrt{2}\sqrt{\eta}\alpha)^2}{1 + \cosh 2 r}\right]}{2 \pi ^2 (1 + \cosh 2 r)^4}\\
    &\bigg(8 \alpha ^2 \eta ^2-4 \eta -4 (\eta -1) \cosh (2 r)+\cosh (4 r)+2 \eta  x_1^2 +4 \sqrt{2} \alpha  \eta ^{3/2} x_1+2 \eta  x_2^2+2 \eta  y_1^2+2 \eta  y_2^2+4 \sqrt{2} \alpha  \eta ^{3/2} y_2+3\bigg).
\end{align}
It is clear to see that despite the contributions in the coefficient term arising from the sent quantum state $\hat \rho$, the leading exponential term describes a predominantly Gaussian distribution centred around approximately $(\sqrt{\eta} \alpha, 0, 0, \sqrt{\eta} \alpha)^T$, i.e. Alice's initial displacement modulated by the effect of the channel. 

We note here that the fact that the state output by the teleporter is equivalent to the original non-displaced state displaced by some attenuated amount is a consequence of the commutativity of the pure-loss channel with the displacement operator, i.e. it is straightforward to see that in general for a thermal-loss channel  $\mathcal{E}[\eta, \e]$  parameterised by loss $\eta$ and thermal noise $\e$
\begin{align}
    \mathcal{E}[\eta, \e] \left(\hat D^\dagger(\bm{x})\hat \rho\hat D^\dagger(\bm{x}) \right) \Longleftrightarrow \hat D(\sqrt{\eta}\bm{x}) \mathcal{E}[\eta, \e] (\hat \rho) \hat D^\dagger(\sqrt{\eta}\bm{x})
\end{align}
This property holds for the `physical channels' most commonly seen when modelling telecommunications protocols, and so we make this assumption when deriving Eq. \eqref{eq:cmqc_channel_effective_1} in the main body of the text. Furthermore, if the channel $\mathcal{E}^N$ obeys this property but is not a lossy bosonic channel, we can write a more general form of Eq. \eqref{eq:cmqc_channel_effective}
\begin{align}
    \mathcal{E}_\mathrm{CMQC}(\hat \rho) &= \hat D(\xi\sqrt{\tau} \bm{\alpha} - g_\alpha \langle \bm{\alpha} \rangle) \Lambda^{\otimes N}[\tau] \left( \mathcal{E}^N \left( \hat \rho \right) \right) \hat D^\dagger(\xi\sqrt{\tau} \bm{\alpha} - g_\alpha \langle \bm{\alpha} \rangle)
\end{align}
and so in the limit of high classical signal we obtain
\begin{align}
    \mathcal{E}_\mathrm{CMQC}(\hat \rho) &\longrightarrow \Lambda^{\otimes N}[\tau] \left( \mathcal{E}^N \left( \hat \rho \right) \right)
\end{align}
and in the limit of high classical signal and high squeezing we obtain 
\begin{align}
    \mathcal{E}_\mathrm{CMQC}(\hat \rho) &\longrightarrow \mathcal{E}^N \left( \hat \rho \right).
\end{align}
Hence the CMQC circuit can be said to more generally reproduce the output of the communications channel regardless of the channel itself, so long as Bob knows the effect of the channel on the displacement and that the channel itself commutes with the displacement operation.

Continuing on, in the instance where Alice instead sends e.g. $(-\alpha, 0, 0, \alpha)^T$, the distribution $P(\bm{x})$ is centred around $(-\sqrt{\eta} \alpha, 0, 0, \sqrt{\eta} \alpha)^T$ corresponding with the symbols $\{\tilde \alpha_2, \tilde \alpha_1 \}$ and so on. Bob can use this to infer which symbols Alice has sent by reasoning that, assuming the size of the classical displacement is large compared to the extent of the quantum state, with a high probability the values of the homodyne measurements will reflect the incoming symbols. 

For each shot, Bob's estimates $\{\tilde \beta_1, \tilde \beta_2\}$ of the sent symbols on each mode respectively are thus computed from $\bm{x}$ via the decoding mechanism for standard coherent-state BPSK \cite{xiong_digital_2006}:
\begin{align}
    \tilde \beta_1 &= \begin{cases}
        \tilde \alpha_1 & x_1 \geq 0 \\
        \tilde \alpha_2 & x_1 < 0 \\
    \end{cases}
    \hspace{2cm} 
    \tilde \beta_2 = \begin{cases}
        \tilde \alpha_1 & y_2 \geq 0 \\
        \tilde \alpha_2 & y_2 < 0 \\
    \end{cases}
    \label{eq:bpsk_decoder}
\end{align}
Of course, the decoding mechanism in Eq. \eqref{eq:bpsk_decoder} is necessarily imperfect on a shot-by-shot basis: depending on the ratio of the extent of the quantum state in the phase space to the size of the classical amplitude, i.e. the signal-to-noise (SNR) ratio, on some occasions the quantum noise associated with the detection will cause the measurement outcome to lie outside the correct region with some small probability $e_S$, called the symbol-error rate (SER). In these cases Bob will estimate the wrong classical symbol from the sent state. For the Bell-state protocol using a single symbol $\{\tilde \alpha_1, \tilde \alpha_1\}$ described here, the SER of each arm follows simply from integrating the probability distribution $P$ over the appropriate decoding regions in Eq. \eqref{eq:bpsk_decoder}:
\begin{align}
    e_S &= 1 - \int \dd y_1 \ \dd x_2 \ \dd y_2 \int_{-\infty}^{0} \ \dd x_1 \ P(\bm{x}) \\
    &= \frac{1}{2} \mathrm{erfc}\left( \frac{\sqrt{\eta} \alpha}{\cosh r} \right)+\frac{\alpha  \eta ^{3/2} \sech^3 r \ e^{-\alpha ^2 \eta  \sech^2(r)}}{4\sqrt{\pi }}.
\end{align}
The bit-error rate (BER) $e_C$ of the protocol is equal to the SER since each arm encodes only a single bit; in addition, the symmetry of the classical encoding guarantees that both arms have the same SER $e_S$, and also that the SER of the protocol remains $e_S$ regardless of the input symbols $\{\tilde \alpha_i, \tilde \alpha_i\}$.

Given that Bob can reconstruct the classical message in this way, he can also predict the mean displacement of his teleported state $\hat \rho_\mathrm{tele}$ via the same method. The final stage of the protocol thus requires Bob to perform the inverse displacement to return the state to its original non-displaced form. However, because Bob cannot exactly estimate the symbols sent by Alice, he cannot exactly reverse the displacement; instead, he can only perform an approximate operation 
\begin{align}
    \hat D(g_\alpha\langle \bm{\alpha} \rangle) &= \bigotimes_i\hat D_{B_i}(g_\alpha\langle \bm{\alpha} \rangle)
\end{align}
where for each mode
\begin{align}
    \hat D_{B_1}(g_\alpha\langle \bm{\alpha} \rangle) &= \begin{cases}
        \hat D_{B_1}[-g_\alpha (+\alpha, 0)^T] & x_1 \geq 0 \\
        \hat D_{B_1}[-g_\alpha (-\alpha, 0)^T] & x_1 < 0 \\
    \end{cases}
    \hspace{1cm}
    \hat D_{B_2}(g_\alpha\langle \bm{\alpha} \rangle) = \begin{cases}
        \hat D_{B_2}[-g_\alpha (0, +\alpha)^T] & y_2 \geq 0 \\
        \hat D_{B_2}[-g_\alpha (0, -\alpha)^T] & y_2 < 0. \\
    \end{cases}
\end{align}
It is clear to see that this operation corresponds to Bob performing the appropriate displacement on each mode to invert the classical decoding he assumes the state to have based on his measurement values $\bm{x}$. The electronic gain $g_\alpha$ is chosen appropriately such that when Bob performs the correct displacement the state is returned to its original mean, modulo any channel effects and the effect of the teleporter; hence we find
\begin{align}
    g_\alpha &= \sqrt{2}\sqrt{\eta}\tanh r \equiv \sqrt{2} \sqrt{\tau \eta}
\end{align}
where $\sqrt{\tau \eta}$ represents the combined attenuation of the classical displacement as a result of the pure-loss communication channels $\Lambda^{\otimes 2}(\eta)$ as well as the effective pure-loss channel $\Lambda^{\otimes 2}(\tau)$ generated by the teleporter.

The final output state, $\hat \rho_\mathrm{out}$, after the re-displacement operation on a shot-by-shot basis is thus given by
\begin{align}
    \hat \rho_\mathrm{out} &= \hat D(g_\alpha\langle \bm{\alpha} \rangle) \hat \rho_\mathrm{tele} \hat D^\dagger(g_\alpha\langle \bm{\alpha} \rangle).
\end{align}
In the Wigner representation, we can write the Wigner function of the state by using Lemma \ref{lem:wignerfunctransform} to implicitly perform the displacement operation:
\begin{align}
    W_{\mathrm{out}}(\bm q) &= \begin{cases}
        W_{\mathrm{tele}}[\bm q - (0,0,0,0,-g_\alpha \alpha, 0, 0, -g_\alpha\alpha)^T] & x_1 \geq 0, y_1 \geq 0 \\ 
        W_{\mathrm{tele}}[\bm q - (0,0,0,0,-g_\alpha\alpha, 0, 0,  g_\alpha\alpha)^T] & x_1 \geq 0, y_1   <  0 \\ 
        W_{\mathrm{tele}}[\bm q - (0,0,0,0, g_\alpha\alpha, 0, 0, -g_\alpha\alpha)^T] & x_1   <  0, y_1 \geq 0 \\ 
        W_{\mathrm{tele}}[\bm q - (0,0,0,0, g_\alpha\alpha, 0, 0,  g_\alpha\alpha)^T] & x_1   <  0, y_1   <  0. \\ 
    \end{cases}
\end{align}
It is important to remember that $\hat \rho_\mathrm{ens}$ depends both implicitly and explicitly on the measurement outcome $\bm{x}$, since we consider only a shot at a time and that the state is not purified on a shot-by-shot basis. It is thus more useful when analysing the protocol to cast Bob's output in terms of the ensemble state, $\hat \rho_\mathrm{ens}$, which is obtained by Bob after asymptotically many rounds of the protocol. In this scenario, the output state is a stochastic mixture of $\hat \rho_\mathrm{out}$ distributing according to the probability density function $P(\bm{x})$:
\begin{align}
    \hat \rho_\mathrm{ens} &= \int \dd^4 \bm{x} \ P(\bm{x}) \hat \rho_\mathrm{out}.
\end{align}
The analytic expression for Wigner function of the ensemble state is thus
\begin{align}
    W_{\mathrm{ens}}(\bm q) = &\int \dd x_2 \dd y_1 \int_{0}^{\infty} \dd x_1 \int_{0}^{\infty} \dd y_2 \ P(\bm{x}) \hat W_\mathrm{tele}[\bm q - (0,0,0,0,-g_\alpha \alpha, 0, 0, -g_\alpha\alpha)^T] \\
    &+\int \dd x_2 \dd y_1 \int_{0}^{\infty} \dd x_1 \int_{-\infty}^{0} \dd y_2 \ P(\bm{x}) \hat W_\mathrm{tele}[\bm q - (0,0,0,0,-g_\alpha \alpha, 0, 0, g_\alpha\alpha)^T] \\
    &+\int \dd x_2 \dd y_1 \int_{-\infty}^{0} \dd x_1 \int_{0}^{\infty} \dd y_2 \ P(\bm{x}) \hat W_\mathrm{tele}[\bm q - (0,0,0,0,g_\alpha \alpha, 0, 0, -g_\alpha\alpha)^T] \\
    &+\int \dd x_2 \dd y_1 \int_{-\infty}^{0} \dd x_1 \int_{-\infty}^{0} \dd y_2 \ P(\bm{x}) \hat W_\mathrm{tele}[\bm q - (0,0,0,0,g_\alpha \alpha, 0, 0, g_\alpha\alpha)^T].
\end{align}

\vspace{1em}
\textit{Violating the CHSH inequality using the CMQC protocol.}{---}An appropriate test for the quality of the Bell state output by the CMQC protocol is to determine whether the state is Bell-violating \cite{bell_einstein_1964}. Violating a Bell inequality necessarily indicates that a state possesses non-classical correlations, and is the strongest possible requirement on the entanglement of a state \cite{wiseman_steering_2007}; in fact, the production of Bell-violating entangled pairs is a prerequisite for the function of protocols that require enhanced quantum correlations, such as device-independent QKD \cite{brukner_bells_2004, masanes_secure_2011}. It is therefore pertinent to determine whether the CMQC protocol is capable of outputting entangled states which can violate the Bell inequality.

The most straightforward Bell test for our purposes is the Clauser-Horne-Shimony-Holt (CHSH) inequality \cite{clauser_proposed_1969}, which states that for a quantum state $\hat \rho$ the expectation value
\begin{align}
    S &= \langle \hat A_0 \otimes \hat B_0 \rangle + \langle \hat A_1 \otimes \hat B_0 \rangle + \langle \hat A_0 \otimes \hat B_1 \rangle - \langle \hat A_1 \otimes \hat B_1 \rangle \label{eq:chsh}
\end{align}
where
\begin{align}
    \hat A_0 = \hat \sigma_Z &\hspace{2cm} \hat B_0 = -\frac{\hat \sigma_Z + \hat \sigma_X}{\sqrt{2}} \\
     \hat A_1 = \hat \sigma_X &\hspace{2cm} \hat B_1 = -\frac{\hat \sigma_Z - \hat \sigma_X}{\sqrt{2}}
\end{align}
is Bell-violating, i.e. the underlying correlations admit only a non-classical model, for $\abs{S} > 2$. The upper bound on the value $\abs{S} \leq 2\sqrt{2}$ is given by Cirel'son \cite{cirelson_quantum_1980}, with the bound being saturated for the maximally-entangled Bell state $\ket{\phi^-}$. 

As before, we compute $S$ using the Wigner representation. The expectation values in Eq. \eqref{eq:chsh} are each given by the integral of the Wigner function of the ensemble state $W_{\mathrm{ens}}$ weighted by the Wigner function of the operators, $W_{\hat A_i \otimes \hat B_i}$; we are thus tasked with determining $W_{\hat A_i \otimes \hat B_i}$, keeping in mind that the Pauli operators act on the logical states of the Bell state $\ket{\phi^-}$ and so must be converted to a physical basis before finding the Wigner function. To begin, we observe that in general for product operators
\begin{align}
    W_{\hat A_i \otimes \hat B_i}(\bm{q}) &= W_{\hat A_i}(\bm{q}_A) W_{\hat B_i}(\bm{q}_B). \label{eq:wignerfunc_product_AB}
\end{align}
Now, in general any operator acting on a logical qubit can be written in terms of a matrix representation
\begin{align}
    \hat O &= \begin{pmatrix}
        a & b \\
        c & d 
    \end{pmatrix}
\end{align}
which admits decomposition in terms of the basis states
\begin{align}
    \hat O \equiv a \ket{0}_L\bra{0}_L + b \ket{0}_L\bra{1}_L + c \ket{1}_L\bra{0}_L + d \ket{1}_L\bra{1}_L
\end{align}
and so has Wigner representation
\begin{align}
    W_{\hat O}(\bm q) &= a W_{\ket{0}_L\bra{0}_L}(\bm q) + b W_{\ket{0}_L\bra{1}_L}(\bm q) + c W_{\ket{1}_L\bra{0}_L}(\bm q) + d W_{\ket{1}_L\bra{1}_L}(\bm q).
\end{align}
This follows simply from the definition of the Wigner function. Recalling that in the dual-rail encoding, our logical states relevant to the output state $\hat \rho_\mathrm{ens}$ are defined for each mode for $X \equiv A,B$ by
\begin{align}
    \ket{0}_L &= \ket{10}_{X_1 X_2} \\
    \ket{1}_L &= \ket{01}_{X_1 X_2}
\end{align}
thus
\begin{align}
    \ket{0}_L\bra{0}_L &= \ket{10}_{X_1 X_2} \bra{10}_{X_1 X_2} = \ket{1}\bra{1}_{X_1} \otimes \ket{0}\bra{0}_{X_2} \\
    \ket{0}_L\bra{1}_L &= \dots
\end{align}
and so on. Using again the fact that the Wigner function of a product operator is the product of the Wigner functions, we can write the Wigner function of the operator $\hat O$ acting on the dual-rail qubit via
\begin{align}
    W_{\hat O}(\bm q) = &a W_{\ket{1}\bra{1}_{X_1}}(\bm q_{X_1})W_{\ket{0}\bra{0}_{X_2}}(\bm q_{X_2}) + b W_{\ket{1}\bra{0}_{X_1}}(\bm q_{X_1})W_{\ket{0}\bra{1}_{X_2}}(\bm q_{X_2}) \notag\\
    &+ c W_{\ket{0}\bra{1}_{X_1}}(\bm q_{X_1})W_{\ket{1}\bra{0}_{X_2}}(\bm q_{X_2}) + d W_{\ket{0}\bra{0}_{X_1}}(\bm q_{X_1})W_{\ket{1}\bra{1}_{X_2}}(\bm q_{X_2}). \label{eq:wignerfunc_of_qubit_operator}
\end{align}
At this stage we can substitute known results for the Wigner function of the single-photon and vacuum states
\begin{align}
    \begin{split}
        W_{\ket{0}\bra{0}}(q,p) &= \frac{e^{-\frac{1}{2}\left( q^2 + p^2 \right)}}{2\pi} \\
        W_{\ket{1}\bra{1}}(q,p) &= \frac{(- 1 + q^2 + p^2)e^{-\frac{1}{2}\left( q^2 + p^2 \right)}}{2\pi}
    \end{split}
    \hspace{1cm}
    \begin{split}
        W_{\ket{0}\bra{1}}(q,p) &= \frac{(-q-ip)e^{-\frac{1}{2}\left( q^2 + p^2 \right)}}{2\pi} \\
        W_{\ket{0}\bra{1}}(q,p) &= \frac{(-q+ip)e^{-\frac{1}{2}\left( q^2 + p^2 \right)}}{2\pi}
    \end{split}
\end{align}
into Eqs. \eqref{eq:wignerfunc_of_qubit_operator} and \eqref{eq:wignerfunc_product_AB} to obtain the Wigner functions of the four CHSH operators:
\begin{align}
    W_{\hat A_0\otimes \hat B_0}(\bm q) &= -\frac{\left(q_{A_1}^2-q_{A_2}^2+p_{A_1}^2-p_{A_2}^2\right) \left[ q_{B_1}^2 + 2 q_{B_1} q_{B_2} - q_{B_2}^2 + p_{B_1}^2 + 2 p_{B_1} p_{B_2} - p_{B_2}^2 \right] e^{-\frac{1}{2} \bm{q}^T\bm{q}}}{16 \sqrt{2} \pi ^4} \\
    W_{\hat A_0\otimes \hat B_1}(\bm q) &= -\frac{\left(q_{A_1}^2-q_{A_2}^2+p_{A_1}^2-p_{A_2}^2\right) \left[ q_{B_1}^2 - 2 q_{B_1} q_{B_2} - q_{B_2}^2 + p_{B_1}^2 - 2 p_{B_1} p_{B_2} - p_{B_2}^2 \right] e^{-\frac{1}{2} \bm{q}^T\bm{q}}}{16 \sqrt{2} \pi ^4} \\
    W_{\hat A_1\otimes \hat B_0}(\bm q) &= -\frac{\left(q_{A_1}q_{A_2} + p_{A_1}p_{A_2}\right) \left[ q_{B_1}^2 + 2 q_{B_1} q_{B_2} - q_{B_2}^2 + p_{B_1}^2 + 2 p_{B_1} p_{B_2} - p_{B_2}^2 \right] e^{-\frac{1}{2} \bm{q}^T\bm{q}}}{8 \sqrt{2} \pi ^4} \\
    W_{\hat A_1\otimes \hat B_1}(\bm q) &= -\frac{\left(q_{A_1}q_{A_2} + p_{A_1}p_{A_2}\right) \left[ q_{B_1}^2 - 2 q_{B_1} q_{B_2} - q_{B_2}^2 + p_{B_1}^2 - 2 p_{B_1} p_{B_2} - p_{B_2}^2 \right] e^{-\frac{1}{2} \bm{q}^T\bm{q}}}{8 \sqrt{2} \pi ^4}.
\end{align}
To evaluate $S$ for the state $\hat \rho_\mathrm{ens}$, we compute
\begin{align}
    S &= \Tr[\hat \rho_\mathrm{ens}(\hat A_0 \otimes \hat B_0)] + \Tr[\hat \rho_\mathrm{ens}(\hat A_0 \otimes \hat B_1)] + \Tr[\hat \rho_\mathrm{ens}(\hat A_1 \otimes \hat B_0)] - \Tr[\hat \rho_\mathrm{ens}(\hat A_1 \otimes \hat B_1)] \\
    &= (4\pi)^4 \int \dd^{8}\bm{q} \ W_{\mathrm{ens}}(\bm{q}) \left[ W_{\hat A_0\otimes \hat B_0}(\bm q) + W_{\hat A_0\otimes \hat B_1}(\bm q) + W_{\hat A_1\otimes \hat B_0}(\bm q) - W_{\hat A_1\otimes \hat B_1}(\bm q) \right].
\end{align}

\textit{Limiting results.}{---}It is useful to perform some tests on the CHSH criterion $S$ to verify the validity of the output and to gain some insight into the function of the protocol. To begin, we determine the limit of $S$ in the case of perfect transmission and no classical communication, i.e. teleportation of one half of the Bell state via a CV teleporter of EPR squeezing $r$:
\begin{align}
    S \underset{\alpha = 0}{\underset{\eta = 1}{\longrightarrow}} 2\sqrt{2} \tanh^2r = 2\sqrt{2} \tau.
\end{align}
Hence we see that the teleporting circuit approaches the upper bound of $S$ for the limit of ideal squeezing $r \longrightarrow \infty$, as the teleporter approaches a pure-loss channel with $\tau = 1$, corresponding with the fact that CV teleportation approaches ideal teleportation for infinite squeezing. 

Second, we consider the case with no classical communication only:
\begin{align}
    S \underset{\alpha = 0}{\longrightarrow} 2\sqrt{2} \eta \tanh^2r = 2\sqrt{2} \tau \eta.
\end{align}
We draw a parallel here between the above result and the result for a dual-rail Bell state simply transmitted through a pair of pure-loss channels of transmissivity $\nu$, which is simple to calculate from the theory above. In the case of ordinary transmission, we find
\begin{align}
    S_\mathrm{pure-loss} &= 2\sqrt{2} \nu
\end{align}
implying that for the pure-loss channel the Bell criterion scales with the transmission $\eta$ and only saturates the bound for $\eta = 1$, i.e. no channel at all. This corresponds to Equation \eqref{eq:cmqc_channel_effective} in the main text, where the total channel describing the protocol is a pure-loss channel of transmissivity $\nu \equiv \tau \eta$. $S_\mathrm{pure-loss}$ is in general an upper bound on $S$; this is reasonable, since the teleportation circuit can only at best reproduce the state $\mathcal{E}(\hat \rho)$ in the limit of perfect classical communication and perfect teleportation.

Next, we consider the limit of the protocol under perfect classical communication, $\alpha \longrightarrow \infty$, where by construction the symbol error rate $e_S$ approaches zero exponentially fast. In this case, Bob can perfectly remove the classical encoding, and so the error contribution on the ensemble state arising from classical mistakes becomes zero. We find
\begin{align}
    \lim_{\alpha \longrightarrow\infty} S &= 2\sqrt{2} \eta \tanh^2 r = 2\sqrt{2} \tau \eta
\end{align}
indicating that the CMQC protocol approaches the equivalent protocol without classical encoding in the limit of a strong enough classical signal. This is a property shared with previous SQCC formulations \cite{zaunders_enhanced_2025} and in essence allows the user to fully separate the classical and quantum data streams. We note as well that the above result approaches $S_\mathrm{pure-loss}$ in the limit of perfect teleportation, and so we can say that the output of the CMQC protocol is channel limited in the case of perfect teleportation and perfect classical communication, i.e. the protocol achieves the best possible result.

Lastly, we consider the interesting case where we assume finite classical communication strength but infinite squeezing. The expectation might be that $S$ approaches $S_\mathrm{pure-loss}$; however, we find
\begin{align} \label{eq:S_limit_high_squeezing}
    \lim_{r \longrightarrow\infty} S &= \frac{\eta}{\sqrt{2}} + \frac{1}{\sqrt{2}}\left[ \eta e^{-8\alpha^2 \eta} - 8 \alpha^2 \eta^2 e^{-8\alpha^2 \eta} + 2\eta (1 - 2\alpha^2 \eta)^2 e^{-4\alpha^2 \eta} \right].
\end{align}
In the limit of perfect classical communication $\alpha \longrightarrow \infty$, we now find that $S \longrightarrow \eta/\sqrt{2}$, i.e. the circuit with perfect teleportation is now unable to violate the Bell inequality. The reason for this discrepancy is that as squeezing grows relative to a finite classical signal $\alpha$, the SNR decreases exponentially fast; this can be seen in the leading exponential term of $P(\bm x)$, which appears like a Gaussian with variance $1 + \cosh 2r$ (corresponding to contributions from the shot noise and from the noise of the EPR state, $V_{B_iB_{i_s}} = \cosh 2r$). These classical errors degrade the output quantum state significantly, even in comparison to the increase in state purity obtained from the improved teleportation. We can therefore conclude that an advantage of the CMQC protocol is that the quality of the output state increases exponentially with the squeezing of the resource EPR state, and so the protocol is experimentally feasible and does not require infinite squeezing; on the other hand, increasing the classical signal strength provides a substantial benefit to performance for very little cost resource-wise. 

\begin{figure}
    \centering
    \includegraphics{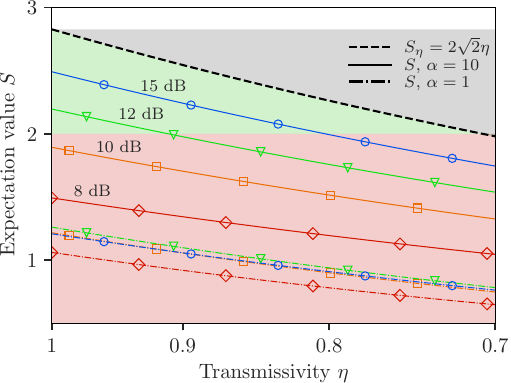}
    \hfill
    \includegraphics{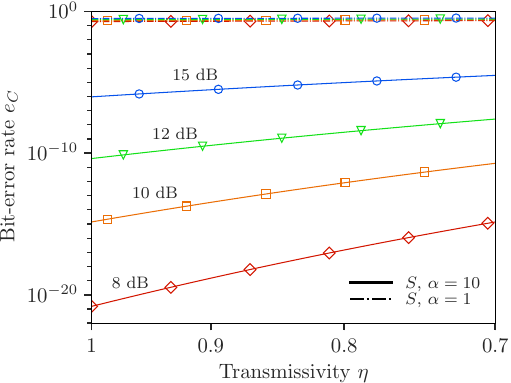}
    \caption{(a) Numerical evaluation of the CHSH criterion $S$ for an instance of the CMQC protocol where the resource state is the dual-rail Bell state $\hat \rho = \ket{\phi^-}\bra{\phi^-}$ and the channels are identical pure-loss channels of transmissivity $\eta$. $S$ is plotted as a function of the loss $\eta$ for the case where the classical signal strength $\alpha = 1$ (dash-dotted lines) and $\alpha = 10$ (solid lines). The squeezing $r$ of the teleporter resource EPR state is varied between multiple values representing feasible experimental parameters (e.g. \cite{takeno_observation_2007, mehmet_observation_2010, stefszky_balanced_2012, eberle_stable_2013, chekhova_bright_2015, vahlbruch_detection_2016}. For sufficiently high signal strength, as teleporter squeezing increases $S$ begins to approach the upper bound $S_\eta = 2\sqrt{2} \eta$, which describes the CHSH criterion for the asymmetric transmission of a Bell state through a pure-loss channel; in contrast, for low signal strength the protocol does not benefit from improved squeezing, with $S$ decreasing above approximately $r = 12$ dB. (b) Bit-error rate of the protocols shown in Fig. \ref{fig:pubfig_bellstate}.a. For high classical signal ($\alpha = 10$), the error rate of the protocol remains essentially negligible, albeit increasing exponentially with squeezing $r$. For low classical signal ($\alpha = 1$) the bit-error rate is effectively saturated regardless of loss or squeezing.}
    \label{fig:pubfig_bellstate}
\end{figure}

Figure \ref{fig:pubfig_bellstate}.a shows $S$ as a function of the channel loss $\eta$ for some experimentally-feasible values of the classical signal strength $\alpha$ and teleportation squeezing $r$. It can be seen that the protocol is capable of violating the CHSH inequality under realistic conditions. As $r \longrightarrow \infty$, $S$ approaches $S_\mathrm{pure-loss}$, i.e. the degree to which the CMQC state violates the CHSH inequality is indistinguishable from direct transmission of $\hat \rho$ over the channel $\Lambda^{\otimes2}[\eta]$. On the other hand, for low classical signal the maximal $S$ is not achieved under the maximum squeezing $r = 15$ dB as a result of the additional quantum noise introduced by the classical bit errors.
Hence the state is actually unable to violate the CHSH inequality even in the limit of perfect teleportation. The corollary of this is that the CMQC protocol can attain nearly-ideal operation without requiring infinite squeezing as long as the classical signal is sufficiently high. Figure \ref{fig:pubfig_bellstate} also proves that the CMQC protocol is capable of transmitting legitimate quantum entanglement at a rate comparable to an equivalent non-classically-signalling protocol.

For completeness, the bit-error rate of the protocol for the conditions given in Figure \ref{fig:pubfig_bellstate}.a is given in Fig. \ref{fig:pubfig_bellstate}.b. For reasonable values of the classical signal strength the bit-error rate becomes essentially negligible, scaling with loss and with signal strength essentially indistinguishably from an ordinary classical communications protocol. It should be noted that while in-principle perfect classical communications can be obtained for any state by increasing $\alpha$ to arbitrarily large levels, the Bell state example is particularly conducive to classical communications because the contribution of the quantum signal to the classical is at most a single photon. However, if one were to use the CMQC protocol to perform e.g. continuous-variable entanglement distribution using an EPR state as the quantum resource $\hat \rho$, then the quantum noise associated with the state then would non-negligibly affect the coupling between the quantum signal and the classical signal and vice versa.

\textit{Post-selected Bell criterion.}{---}Finally, we analyse the protocol in terms of the post-selected CHSH criterion. Post-selection is a technique used in practical Bell-violation experiments \cite{branciard_detection_2011} to overcome the fundamental rate-loss scaling of the loophole-free Bell test; namely, the loophole-free CHSH criterion
\begin{align}
    S_\mathrm{pure-loss} &= 2\sqrt{2} \nu
\end{align}
cannot exhibit Bell-violating correlations below a transmission of $\nu = \frac{1}{\sqrt{2}} \simeq 0.71$. This is not necessarily a limitation of the physicality of the system, since we know the state still represents a quantum correlation, but rather an indication of the strictness of loophole-free Bell tests. The stringency in this case comes from the requirement of fair sampling \cite{garg_detector_1987, giustina_bell_2013, gisin_local_1999}, where we must consider all measurements when calculating the CHSH criterion $S$ such that the measurements construe a fair and representative sample of the actual quantum state. 

Abandoning this requirement by discarding unfavourable measurement results allows us to estimate $S$ in a way that is arguably more accurate to the original state; the archetypical example is given by Branciard \cite{branciard_detection_2011} for the case of a Bell state measured by Alice and Bob with inefficient detectors. Assume that Alice and Bob wish to measure whether an entangled state is Bell-violating with faulty detectors, such that each detector has probability $\nu$ of incorrectly `losing' a photon. In the cases where a photon is lost, Alice and Bob record the measurement result \{0,0\}. Knowing that this result can only occur as a result of detector inefficiency, Alice and Bob discard it. Their measured correlations in the postselected case are given by an application of Bayes' theorem \cite{branciard_detection_2011}:
\begin{align}
    S_p &= \frac{S}{p_s}
\end{align}
where $S$ is the original estimate of the correlations including both conclusive events (measurement outcomes compatible with successful detection) and nonconclusive events (measurement outcomes incompatible with successful detection). $p_s$ the probability of obtaining a conclusive event. 

We motivate the existence of postselection by assuming that we now transmit a \textit{logical} Bell state, i.e. one half of a Bell pair encoded according to some loss-tolerant error correction scheme such as a parity code \cite{hayes_loss-tolerant_2008, muralidharan_ultrafast_2014, ewert_ultrafast_2016, ewert_ultrafast_2017}, and measure the error syndrome of the multi-mode logical state after completing the teleportation. Any event where an error is heralded by the syndrome is discarded. Consider again our simple example of a dual-rail Bell state, where one half is retained and the other transmitted over a pair of pure-loss channels $\Lambda[\nu]$. It is clear to see that in this configuration, if Bob heralds an error, he knows the photon has been lost to the environment; he therefore labels the event as nonconclusive and discards the event. We model this by defining the probability of obtaining a conclusive event as equivalent to one where he measures exclusively one photon and vacuum on alternate rails, i.e. the photon actually appears at the station. The probability of this is
\begin{align}
    p_s^\mathrm{pure-loss} &= \Tr[\mathcal{E}(\hat \rho) \hat \Pi]
\end{align}
for 
\begin{align}
    \hat \Pi = \ket{0}\bra{0} \otimes \ket{1}\bra{1} + \ket{0}\bra{0} \otimes \ket{1}\bra{1}.
\end{align}
In the Wigner basis, we find
\begin{align}
    p_s^\mathrm{pure-loss} &= \int \dd^4\bm{q} \ W_{\mathcal{E}(\hat \rho)}(\bm{q}) W_{\hat \Pi}(\bm{q}) \\
    &= \int \dd^4\bm{q} \ 
    W_{\mathcal{E}(\hat \rho)}(\bm{q}) \left[ W_{\ket{0}\bra{0}}(\bm{q}_{B_1})W_{\ket{1}\bra{1}}(\bm{q}_{B_2}) + W_{\ket{1}\bra{1}}(\bm{q}_{B_1})W_{\ket{0}\bra{0}}(\bm{q}_{B_2}) \right] \\
    &= \nu.
\end{align}
Thus $S_p^\mathrm{pure-loss} = 2\sqrt{2}$. Since the rate $R$ at which Bob obtains conclusive events is equal to the probability of success, i.e. $R = \eta$, Bob can obtain perfectly correlated statistics at a rate that scales proportionally to the losses of the two channels.

The same logic applies to the state output by the CMQC protocol when transmitting one half of a Bell state, which we have derived above. Now, the probability of success is instead given by
\begin{align}
    p_s &= \int \dd^4\bm{q} \ W_\mathrm{ens}(\bm{q}) W_{\hat \Pi}(\bm{q}).
\end{align}

In the limit of high classical signal, we find
\begin{align}
    \lim_{\alpha \longrightarrow \infty} p_s &= \eta \tanh^2 r = \tau \eta
\end{align}
i.e. approaching the non-CMQC result $S_p^\mathrm{pure-loss} = \eta$ as the EPR squeezing increases, which is expected. This implies the postselected CHSH criterion $S_p$ in the limit of large classical signal becomes
\begin{align}
    \lim_{\alpha \longrightarrow \infty} S_p &= \lim_{\alpha \longrightarrow \infty} \frac{S}{p_s} = \frac{{\lim_{\alpha \longrightarrow \infty}} S}{\lim_{\alpha \longrightarrow \infty} p_s} = \frac{2\sqrt{2}\eta \tanh^2r}{\eta \tanh^2r} = 2\sqrt{2} 
\end{align}
which is commensurate with the protocol reducing to a pure-loss channel of transmissivity $\tau \eta$ in the limit of perfect classical signalling.

In the limit of finite classical signal and high squeezing, we find
\begin{align}
    \lim_{r\longrightarrow\infty} p_s = \frac{1}{4} + \frac{1}{4} \eta  e^{-8 \alpha ^2 \eta } \left(32 \alpha ^4 \eta ^2+\alpha ^2 (8-16 \eta )+2 e^{4 \alpha ^2 \eta } \left(8 \alpha ^4 \eta ^2+\alpha ^2 (4-8 \eta )+1\right)+1\right).
\end{align}
Thus if $\alpha \longrightarrow \infty$, $p_s \longrightarrow 1/4$. This corresponds to the fact that at high squeezing, the classical SNR of the signal is effectively 0: Bob therefore has only a 25\% chance of guessing the correct displacement and returning the state to a quasi-Bell state. All other guesses produce a state with a large net displacement, which is then discarded upon measurement since they nearly always output a photon measurement greater than 1. In this case the postselected Bell criterion becomes $S_p \longrightarrow (\eta/\sqrt{2}) / (1/4) = 2\sqrt{2} \eta$, i.e. still capable of violating the CHSH inequality but only at extremely low losses.

For completeness, we present the rates (Fig. \ref{fig:supfig_bellstate_postselected}.a) and optimal values of $r$ (Fig. \ref{fig:supfig_bellstate_postselected}.b) corresponding to the curves shown in Figure \ref{fig:pubfig_bellstate_postselected} of the main text.

\begin{figure*}[!htb]
    \centering
    \includegraphics{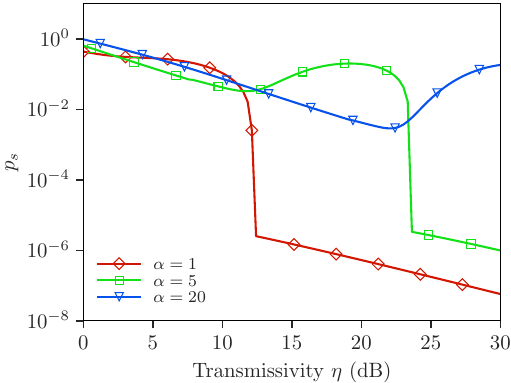}
    \includegraphics{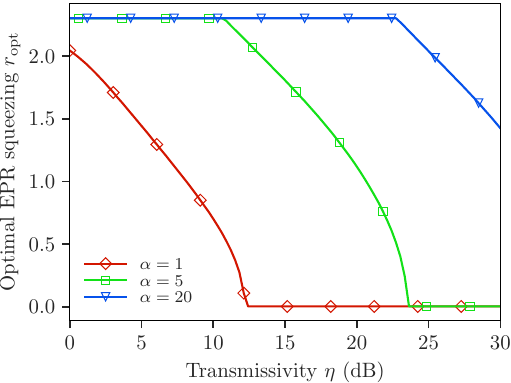}
    \caption{Success probabilities (left) and optimal rates (right) for the postselected protocols shown in Fig. \ref{fig:pubfig_bellstate_postselected} of the main text. The rate at which Bob obtains the correct measurement results with which to evaluate $S$ is given trivially by the probability of success $p_s$. For each loss $\eta$, the postselected quantity $S_p = S/p_s$ is optimised over the two-mode squeezing of Bob's EPR resource $r$ between $r = 0$ and 20 dB ($r \in [0, 2.3]$). The optimal strategy for protocols with a high classical signal $\alpha$ and so high classical SNR is the obvious one, i.e. $r \longrightarrow \infty$. This helps minimise the errors arising from imperfect teleportation so long as classical errors remain low. On the other hand, for low classical signal (such as when $\alpha = 1$ or in the higher-loss regime) classical errors dominate and so it becomes advantageous to reduce the level of squeezing such that the SNR of the protocol is increased.}
    \label{fig:supfig_bellstate_postselected}
\end{figure*}

Finally, we note that a corollary of postselection is that arbitrarily postselecting data in Bell experiments explicitly cannot lead to certified nonlocal correlations \cite{pearle_hidden-variable_1970, gisin_local_1999}. This detection loophole means postselected Bell experiments are not useful for testing the validity of quantum mechanics, unless certain detection-loophole-compatible Bell inequalities are used (see e.g. \cite{giustina_bell_2013, eberhard_background_1993, garg_detector_1987}). On the other hand, postselected Bell violations are perfectly fine for characterising the entanglement of a state which we already know to be quantum in nature, which is how it is employed in this paper.

\end{widetext}

\end{document}